\documentclass[preprint]{revtex4}
\usepackage{epsfig}
\begin{document}
%\draft
\title{Relation between CPT Violation in Neutrino masses and mixings}
\author
{Bipin Singh Koranga$^a$\footnote{bipiniitb@rediffmail.com}, 
Mohan Narayan$^b$\footnote{mohan@udct.org} and 
S. Uma Sankar$^c$\footnote{uma@phy.iitb.ac.in}}
\affiliation
{$^a$Department of Physics, Kirori Mal College  
(University of Delhi), Delhi 110 007. \\
$^b$Department of Physics, 
Institute of Chemical Technology, Mumbai 400 019\\
$^c$Department of Physics, Indian Institute of Technology Bombay,
Mumbai 400 076.}

\begin{abstract}
The neutrino parameters determined from the solar neutrino
data and the anti-neutrino parameters determined from KamLAND 
reactor experiment are in good agreement with each other. However,
the best fit points of the two sets differ from each other 
by about $10^{-5}$ eV$^2$ in mass-square differenc and by 
about $2^\circ$ in the mixing angle. Future solar neutrino  
and reactor anti-neutrino experiments are likely to reduce 
the uncertainties in these measurements. This, in turn, can
lead to a signal for CPT violation in terms a non-zero difference
between neutrino and anti-neutrino parameters. In this paper, 
we propose a CPT violating mass matrix which can give rise to 
the above differences in both mass-squared difference and mixing 
angle and study the constraints imposed by the data on the 
parameters of the mass matrix.
\end{abstract}

\maketitle

\section{Introduction}

KamLAND experiment has established the distortion due to 
oscillations in the anti-neutrino spectrum from reactors 
and determined the corresponding mass-square difference, 
$\overline{\Delta}_{21}$,
to a great precision \cite{klspec,kl2008}. 
At present there is good agreement between $\overline{\Delta}_{21}$
and the mass-squared difference of the neutrinos, $\Delta_{21}$,
determined from the analysis of solar neutrino data
\cite{bahcall04,fogli2006}. However, the best-fit values of the two 
$\Delta$s differ from each other by about $10^{-5}$ eV$^2$. 
Also, the best-fit mixing angles differ from each other by
$2$ to $3$ degrees. 
Together, solar and KamLAND data impose the constraint $|\Delta_{21} 
-\overline{\Delta}_{21}| \leq 1.1 \times 10^{-4}$ eV$^2$
\cite{degouvea}. Future reactor 
experiments, located at a distance of about 70 Km from the 
source so that the oscillation minimum coincides with spectral 
maximum, are expected to improve the precision of anti-neutrino
parameters even further \cite{srubabati}. 
Similarly future solar neutrino experiments \cite{vogelaar}, 
are expected to improve the accuracy of neutrino parameters. 
If these future experiments confirm the present trend in the 
difference between neutrino and anti-neutrino parameters, then 
CPT violation in the neutrino sector becomes an exciting 
possibility \cite{greenberg,kostelecky}. 

When it comes to the larger mass-squared difference, the atmospheric
neutrino data prefers equal values for the neutrino
and anti-neutrino mass-squared differences, though 
the uncertainties do allow large CPT violating effects 
\cite{skadhauge}. MINOS experiment is expected to measure 
the disappearance probability of muon anti-neutrinos with
good precision in near future. If there is any difference between 
$P(\nu_\mu \to \nu_\mu)$ and $P(\bar{\nu}_\mu \to \bar{\nu}_\mu)$,
it will be a signal for CPT violation in the larger mass-squared
difference also \cite{barenlykken}.

We assume that the main part of neutrino mass matrix is
CPT conserving and it arises due to dynamics at an energy scale 
much below the scale at which CPT violation occurs.
We further assume that CPT violation from a high scale,
leads to an addition to the neutrino mass matrix of the form
\begin{equation}
{\cal M_{\rm CPT}} = \mu \lambda_{\alpha \beta}
\label{cptvm}
\end{equation}
where $\alpha$ and $\beta$ are flavour indices and $\mu$ is 
a parameter with dimensions of mass. 
As we show in the next section, to reproduce the allowed differences 
between neutrino and anti-neutrino parameters, we require 
the scale of $\mu$ to be $\sim 10^{-6} \ {\rm eV}$. 
With this as input, we calculate the difference in the mass-squared 
differences and mixing angles for the neutrinos and anti-neutrinos. 

Quantum gravity effects can lead to CPT violation. The leading effective 
operators of quantum gravity are suppressed as the inverse of the
Planck mass $M_{Pl}$. Such operators can give rise to CPT
violating mass $\sim v^2/M_{Pl}$, where $v$ is a low
energy VEV of the quantum gravity model. If $v = 174$ GeV,
the electroweak symmetry breaking scale, then we
obtain $\mu \sim 10^{-6}$ eV.   
The sign of the additional mass matrix for anti-neutrinos 
will have the opposite sign \cite{barenboim}. 

In eq.~(\ref{cptvm}), $\lambda_{\alpha \beta}$ is a $3 \times 3$ 
matrix in flavour space. Quantum gravity effects
are not sensitive to flavour. Hence it is expected that 
every term in the matrix $\lambda_{\alpha \beta}$ is independent 
of both $\alpha$ and $\beta$. We take this matrix to be of the form 
$\lambda_{\alpha \beta} = 1$ for all $\alpha$ and $\beta$. 
In this case, the CPT violating part of the neutrino mass matrix is 
of the form:
\begin{equation}
\mu \textstyle
\left( \begin{array}{lcr}
  1 & \!1\! & 1 \\[-.5ex]
  1 & \!1\! & 1 \\[-.5ex]
  1 & \!1\! & 1 
\end{array} \right). 
\label{textm}
\end{equation}
In our calculations, we take eq.~(\ref{textm}) as a perturbation to the 
main part of the neutrino mass matrix. The pattern of CPT violation
in neutrino and anti-neutrino mass-squared difference due to the above
matrix was analyzed in \cite{cpt2006}. Here we consider the constraints
from data on the CPT violating parameters in mass-squared differences
and mixing angles. 

\section{Calculation}

We assume that the CPT conserving part of the light neutrino mass 
has real and non negative eigenvalues $M_i$. In the mass eigenbasis, 
this matrix appears as $M=diag(M_{1},M_{2},M_{3})$. We treat $M$ 
as the unperturbed ($0^{th}-order$) mass matrix. Denoting the 
corresponding neutrino mixing matrix by $U$, we obtain the $0^{th}-order$ 
mass matrix $\mathbf{M}$ in flavour space as 
\begin{equation}
\mathbf{M}=U^{*}MU^{\dagger}.
\label{defM}
\end{equation}
Explicitly, the matrix $U$ has the form
\begin{equation}
U=\left(\begin{array}{ccc}
U_{e1} & U_{e2} & U_{e3}\\
U_{\mu1} & U_{\mu2} & U_{\mu3}\\
U_{\tau1} & U_{\tau2} & U_{\tau3}\end{array}\right),
\end{equation}
where the nine elements are functions of three mixing angles and
six phases. In terms of the above elements, the mixing angles are 
defined by             
\begin{eqnarray}
\left|\frac{U_{e2}}{U{}_{e1}}\right| & = & \tan\theta_{12},  \label{def12} \\
\left|\frac{U_{\mu3}}{U_{\tau3}}\right| & = & \tan\theta_{23}, \label{def23} \\
|U_{e3}| & = & \sin\theta_{13}. \label{def13}
\end{eqnarray}
In terms of the above mixing angles, the MNS matrix is written
as
\begin{equation}
U=diag(e^{if_1},\,\, e^{if_2},\,\, e^{if_3})R(\theta_{23})\Delta \,\, R(\theta_{13})
\Delta^{*} \, R(\theta_{12})diag(e^{ia_1},e^{ia_2},1).
\end{equation}
The matrix $\Delta=diag(e^{\frac{i\delta}{2}},1,e^{\frac{-i\delta}{2}})$
contains the Dirac phase $\delta.$ This phase leads to CP violation
in neutrino oscillations. $a_1$ and $a_2$ are the so called Majorana
phases, which affect the neutrinoless double beta decay. $f_1, f_2$
and $f_3$ are usually absorbed as a part of the definition of the
charged lepton fields. It is possible to rotate these phases away,
if the mass matrix in eq.~(\ref{defM}) is the complete mass matrix. However,
since we are going to add another contribution to this mass matrix,
these phases of the zeroth order mass matrix can have an impact on
the complete mass matrix and thus must be retained. By the same token,
the Majorana phases which are usually redundant for oscillations have
a dynamical role to play now.

Given the above $0$th order (CPT conserving) neutrino
mass matrix, we now add the CPT violating mass matrix to
it. Thus the complete light neutrino mass matrix contains
both CPT conserving and CPT violating terms. Given that
$\mu$ is much smaller than the light neutrino mass scale
(which should be greater than $\sqrt{\Delta_{atm}}$ eV),
we can treat the CPT violating mass matrix to be a perturbation
of the CPT conserving mass matrix. 
The complete neutrino mass matrix in flavour space is 
\begin{equation}
\mathbf{M}\rightarrow\mathbf{M^{'}=M}+\mu\lambda,
\end{equation}
We assume that the symmetries inherent in ${\bf M}$ lead to tribimaximal 
mixing. But $\mu \lambda$ breaks these symmetries. And hence the
mixing angles given by the total mass matrix ${\bf M'}$ will not be
tribimaximal. Below we compute the deviations from tribimaximality
induced by $\mu \lambda$ as well as the differences in mass-squared
splittings. Note that these deviations will be equal and opposite
for neutrinos and anti-neutrinos because $\mu \lambda$ is CPT violating
and is assumed to have opposite signs for particle and anti-particle.

The perturbation formalism, by which the above computation can be done,
was first developed in ref. \cite{vmb}. Here we briefly recall the main
features for completeness. The matrix relevant for oscillation physics
is the following hermitian matrix
\begin{equation}
\mathbf{M^{'^{\dagger}}M^{'}=}(\mathbf{M}+\mu\lambda)^{\dagger}
(\mathbf{M}+\mu\lambda).
\end{equation}
To the first order in the small parameter $\mu$, the above matrix is
\begin{equation}
\mathbf{M}^{\dagger}\mathbf{M}+\mu\lambda^{\dagger}\mathbf{M}
+\mathbf{M}^{\dagger}\mu\lambda.
\end{equation}
This hermitian matrix is diagonalized by a new unitary matrix $U^{'}$.
The corresponding diagonal matrix $M^{'^{2}}$, correct to first order
in $\mu$, is related to the above matrix by $U^{'}M^{'^{2}}U^{'^{\dagger}}$.
Rewriting $\mathbf{M}$ in the above expression in terms of the diagonal
matrix $M$ we get
\begin{equation}
U^{'}M^{'^{2}}U^{'^{\dagger}}=U(M^{2}+m^{\dagger}M+Mm)U^{\dagger},
\label{defup}
\end{equation}
where
\begin{equation}
m=\mu U^{T}\lambda U.
\label{defm}
\end{equation}
Here $M$ and $M^{'}$ are the diagonal matrices with neutrino masses
correct to $0^{th}$and $1^{st}$order in $\mu$. It is clear
from eq.~(\ref{defup}) that the new mixing matrix can be written as:
\begin{equation}
U^{'}=U(1+i\delta\Theta),
\label{defTheta}
\end{equation}
where $\delta\Theta$ is a hermitian matrix that occurs to first order in
$\mu$. Oscillation physics is unchanged under the transformation
$U\rightarrow UP$, where $P$ is a diagonal phase matrix. We can use
this invariance to set the diagonal elements of the matrix $\delta\Theta$
to be zero.

From eq.~(\ref{defup}) we obtain
\begin{equation}
M^{2}+m^{\dagger}M+Mm=M^{'^{'2}}+[i\delta\Theta,M^{'^{2}}].
\end{equation}
Therefore to first order in $\mu$, the mass squared difference
$\Delta M_{ij}^{2}=M_{i}^{2}-M_{j}^{2}$ get modified as
\begin{equation}
\Delta M_{ij}^{'^{2}}=\Delta M_{ij}^{2}+2(M_{i}Re[m_{ii}]-M_{j}Re[m_{jj}]).
\label{moddelta}
\end{equation}
The non-diagonal elements of $\delta\Theta$ are given by
\begin{equation}
(\delta\Theta)_{ij}=\frac{iRe(m_{ij})(M_{i}+M_{j})}{\Delta M_{ij}^{'^{2}}}
-\frac{Im(m_{ij})(M_{i}-M_{j})}{\Delta M_{ij}^{'^{2}}},
\label{exptheta}
\end{equation}
from which the changes in the mixing matrix can be computed 
by substituting $\delta \Theta$ in eq.~(\ref{defTheta}).

The changes induced by the small parameter $m$ are all 
proportional to the neutrino mass eigenvalues. They will
have their largest values in the case of degenerate masses. 
Hence we assume degenerate neutrino masses $M_i \simeq M$ 
from hereon. In the expression for $(\delta \Theta)_{ij}$ in 
Eq.~(\ref{exptheta}), the second term is utterly negligible 
compared to the first, if we use degenerate masses. Thus we
get a greatly simplified expression \cite{th12dev}
\begin{equation}
(\delta\Theta)_{ij}=\frac{2iMRe(m_{ij})}{\Delta M^2_{ij}},
\label{deltheta}
\end{equation}
where we have substituted the $0$th order mas-square
difference in the denominator because the numerator already
contains a factor of $m$. From Eq.~(\ref{deltheta}), 
it is trivial to see that $(\delta \Theta)_{12}$, whose 
expression contains $\Delta_{21}$ in the denominator, 
is the largest among the $(\delta \Theta)_{ij}$.

Given the form of $(\delta \Theta)_{ij}$, the elements of
the modified mixing matrix can be obtained as
\begin{equation}
U_{\alpha j}^{'} 
=U_{\alpha j} + \delta U_{\alpha j}  
=U_{\alpha j} + 
i \sum_{i=1}^3 U_{\alpha i}(\delta\Theta)_{ij}.
\label{deltaUalpj}
\end{equation}
Knowing $U_{\alpha j}^{'}$, we can define the modified
mixing angles $\theta_{ij}^{'}$ in analology to the 
three equations given in Eqs.~(\ref{def12})-(\ref{def13}).
To compute $\theta_{ij}^{'}$, we first need to compute
the changes in the five matrix elements $\delta U_{ej} \ 
(j=1,2,3)$ and $\delta U_{\alpha 3} \ (\alpha = e, \mu, \tau)$.
Given that $(\delta \Theta)_{13}, (\delta \Theta)_{23} \ll (\delta 
\Theta)_{12}$, we can easily show that the changes in $\theta_{13}$
and $\theta_{23}$ are very small. To obtain $\theta_{12}^{'}$,
we need to evaluate the \cite{th12dev}
\begin{eqnarray}
\delta U_{e1} & = & - U_{e2} \frac{Re(m_{12})}{M_2 - M_1},
\label{deltaUe1} \\
\delta U_{e2} & = &  U_{e1} \frac{Re(m_{12})}{M_2 - M_1},
\label{deltaUe2}
\end{eqnarray}
For later convenience we define the complex numbers $z_{i}
=U_{ei}+U_{\mu i}+U_{\tau i}$, where $U_{\alpha i}$ are, 
in general, functions of all six phases.

In terms of the modified mixing matrix elements, $\theta_{12}^{'}$  
is defined as 
\begin{equation}
\tan\theta_{12}^{'}=\left|\frac{U_{e2}^{`}}{U_{e1}^{'}}\right|
\label{th12p}
\end{equation}
Substituting the expressions from eqs.~(\ref{deltaUalpj})-(\ref{deltaUe2})
in eq.~(\ref{th12p}), we get
\begin{eqnarray}
\tan\theta_{12}^{'} & = & \tan\theta_{12} + 2 \frac{\mu M}{\Delta M^2_{21}}
\frac{|z_1||z_2|}{\cos^2 \theta_{12}}\cos(a_1+a_2)\cos(a_1-a_2) \nonumber \\
& = & \tan \theta_{12} + \varepsilon_\theta
\label{th12pexp}
\end{eqnarray}
%From the above expression, we
%notice that the change vanishes if the Majorana phases $a_1,a_2$ satisfy
%the conditions $a_1+a_2=90^\circ~{\rm or}~270^\circ$ or
%$a_1-a_2=90^\circ~{\rm or}~270^\circ$. This is reflected in the plot
%of $\delta \theta_{12} = \theta_{12}^{'} - \theta_{12}$, given in
%fig.~1. The symmetry of this figure under the transformations
%$a_1 \to 180^\circ - a_1$ and $a_2 \to 180^\circ - a_2$ is also explained
%by the above equation.
The modified solar mass-square difference is given by 
\begin{eqnarray}
\Delta M^{'2}_{21} & = & \Delta M^2_{21} + 2 \mu M \left[ |z_2|^2 \cos(2a_2)
- |z_1|^2 \cos(2a_1)\right] \nonumber \\
& = & \Delta M^2_{21} + \varepsilon_\Delta
\label{deltapr}
\end{eqnarray}  
Eqs.~(\ref{th12pexp}) and~(\ref{deltapr}) give the modified mixing 
angle and mass-squared difference for neutrinos. The corresponding 
quantities for anti-neutrinos can simply be obtained by $\mu \to -\mu$.
Thus we have 
\begin{eqnarray}
\overline{\Delta} M^{'2}_{21} & = & \overline{\Delta} M^2_{21} - 
\varepsilon_\Delta
= \Delta M^2_{21} - \varepsilon_\Delta \nonumber \\
\tan \overline{\theta}^{'}_{12} & = & \tan \overline{\theta}_{12} - 
\varepsilon_\theta
= \tan \theta_{12} - \varepsilon_\theta
\label{nubarch}
\end{eqnarray}  
Note that the change in the mixing angle and the change in the 
mass-square difference have very different dependence on the 
Majorana phases $a_1$ and $a_2$. Therefore it will be straitforward to
satisfy the experimental constraints for some combination of these
two phases. 

\section{Results}

As mentioned in the introduction, we assume that the symmetries
of the $0$th order neutrino mass matrix lead to tribimaximal 
mixing with $\theta_{12} \simeq 35.2^\circ, \theta_{13} = 0$
and $\theta_{23} = 45^\circ$. The best fit value for the reactor
anti-neutrino mixing angle, from KamLAND, is $\overline{\theta}_{12}^{'}
= 36.8^\circ$ \cite{kl2008}. For solar neutrinos it is 
$\theta_{12}^{'} = 32.6^\circ$ \cite{bahcall04,schwetzvalle}.
Thus we see that the anti-neutrino mixing angle is $1.6^\circ$ more
than the tribimaximal prediction whereas the neutrino mixing angle 
is $2.6^\circ$ below the prediction. The differences between the
best fits and the tribimaximal prediction are not equal and opposite. 
But, within the experimental uncertainties, they can be taken to be 
$2^\circ$. It is possible that the shifts in the mixing angles  
for neutrinos and anti-neutrinos are not equal. To explain such shifts, 
we need to invoke, along with CPT violating high scale physics, 
contributions from CPT conserving high scale physics, such as planck 
scale effects \cite{th12dev}. The best fit value of reactor 
anti-neutrino mass-squared difference $\overline{\Delta} M^{2'}_{21} = 
8 \times 10^{-5}$ eV$^2$ \cite{kl2008} and for solar neutrinos, 
the best fit value of the mass-squared difference is 
$\Delta M^{2'}_{12} = 6 \times 10^{-5}$ eV$^2$ \cite{bahcall04,schwetzvalle}.

From eqs.~(\ref{th12pexp}),~(\ref{deltapr}) and~(\ref{nubarch}), 
we find $\Delta M^{2'}_{12} - \overline{\Delta} M^{2'}_{21} = 
2 \varepsilon_\Delta
= - 2 \times 10^{-5}$ eV$^2$ and $\theta_{12}^{'} - \overline{\theta}_{12}^{'}
= 2 \varepsilon_\theta = - 4^\circ$. First we explore the following
question: For what values of the CPT violating parameter $\mu$,
is there an agreement between the data and the hypothesis of 
CPT violating neutrino masses? In the introduction, we argued that
$\mu \sim 10^{-6}$ eV. Here we take $\mu = p \times 10^{-6}$ eV,
where $p$ is a number between $1$ to $10$, and derive the constraints
the data imposes on $p$. 
We take degenerate neutrino masses for light neutrinos
$M_i = 2$ eV, which is the upper limit coming from tritium beta decay 
\cite{mainz} and the neutrino mixing angles to be tribimaximal ones. 
Since $\theta_{13}=0$ in this case, the Dirac phase $\delta$ can be set 
to zero without loss of generality.
The zeroth order value of the smaller mass-square difference $\Delta 
M^2_{21}$ is set to $7 \times 10^{-5}$ eV$^2$, which is the average
of the neutrino and anti-neutrino mass-squared difference. The phases
$f_i$ are set to zero.

With these input values, the expressions for $\varepsilon_\theta$
and $\varepsilon_\Delta$ become
\begin{eqnarray}
2 \varepsilon_\theta & = & 0.04 \ p \left[ \cos(2a_2) + \cos(2a_1) \right] 
= - 4*\pi/180 \nonumber \\
2 \varepsilon_\Delta & = & 8p \times 10^{-6} \left[ \frac{1}{3} \cos(2a_2)
- \frac{2}{3} \cos(2a_1) \right] = 2 \times 10^{-5}. 
\end{eqnarray}
Simplifying these equations, we get the following two conditions 
on $a_1$ and $a_2$
\begin{eqnarray}
\left[ \cos(2a_2) + \cos(2a_1) \right] & = & - 1.75/p \nonumber \\
\left[ \cos(2a_2) - 2 \cos(2a_1) \right] & = & 7.5/p.
\label{numconst}
\end{eqnarray}
Solving these two equations and imposing the condition
that $-1 \leq \cos(2a_1),\cos(2a_2) \leq 1$, gives us
the lower limit $p > 3$. 
For $p=4$, eq.~(\ref{numconst}) gives $a_1=-70^\circ$ and $a_2=35^\circ$.
For $p=6$, these values change to $a_1=-60^\circ$ and $a_2=39^\circ$.
Note that the Majorana phases $a_1$ and $a_2$ should necessarily be
non-zero to satisfy the two constraints in eq.~(\ref{numconst}).

\section{Conclusions}
Both solar and reactor data are well explained by neutrino
oscillations. Fit to solar data give a large region for the neutrino 
mass squared difference in the two flavor parameter space. The fit to 
reactor data however gives a very strongly constrained anti-neutrino
mass squared difference. The best fits of the two mass squared differences
are appreciably different from each other. Further improvement in KamLAND
systematics and future solar neutrino data may further strengthen this 
discrepancy, thus giving a signal for CPT violation.
We have demonstrated that flavour blind CPT violating neutrino 
masses from Planck scale physics can nicely accomodate this discrepancy, 
provided the Majorana phases of the neutrino
mass matrix are appreciably large. 
This effect is crucially dependent on the neutrino
mass spectrum and gives rise to observable difference between 
$\Delta_{21}$ and $\overline{\Delta}_{21}$ only for a degenerate neutrino 
mass spectrum with $m_\nu \simeq 2$ eV,
which is the largest allowed value from tritium beta decay data.
The low value of the common mass implied by the WMAP bound
\cite{tegmark} leads to  negligible difference between $\Delta_{21}$ and 
$\overline{\Delta}_{21}$. This can however be
compensated for by considering a slightly lower scale for the
flavour blind CPT violating mass terms rather than the usual Planck scale. 

As we discussed in section II, the difference between $\Delta_{31}'$
and $\overline{\Delta}_{31}'$  is negligible in this scenario if $\mu
\sim 10^{-6}$ eV. If MINOS experiment were to observe a signal for
CPT violation \cite{barenlykken}, the above difference should be of order
$10^{-3}$ eV$^2$. Accounting for such a large CPT violation in the
current scenario requires the CPT violating mass parameter to be of 
the order of $10^{-3}$ eV. To obtain such a large value, the scale of
CPT violating physics has to be three orders below the Planck scale.
The flavour matrix $\lambda_{\alpha \beta}$, can not be flavour
blind because it would lead to an unacceptably large CPT violation 
for $\Delta_{21}$. Hence an appropriate texture should be imposed on
$\lambda_{\alpha \beta}$.

\underline{\it Acknowledgement:} We would like to thank Francesco
Vissani for a critical reading of the manuscript.

\end{document}